\documentclass[twocolumn,showpacs,preprintnumbers,amsmath,amssymb]{revtex4}
\usepackage{graphicx}
\usepackage{color}
\usepackage{dcolumn}
\usepackage{bm}

\begin{document}
%\title{Normal and abnormal states of oceanic surface waves}
%\title{Coherent dynamics of oceanic surface waves}
%\title{Strongly  coherent  dynamics of stochastic waves causes extreme sea states}
%\title{Strongly  coherent  dynamics  of stochastic oceanic waves causes rogue waves}
%\title{Strongly  coherent  dynamics of stochastic waves causes abnormal sea states}
\title{Effects of coherent  dynamics of stochastic deep-water waves}
\author{A.V. Slunyaev$^{1,2,3}$}
%\thanks{slunyaev@hydro.appl.sci-nnov.ru}
%\author{E.N. Pelinovsky$^{1,2}$}

\affiliation{$^1$National Research University-Higher School of Economics, 25 B. Pechorskaya Street, N. Novgorod 603950, Russia \\
$^2$Institute of Applied Physics, 46 Ulyanova Street, N. Novgorod 603950, Russia \\
$^3$Nizhny Novgorod State Technical University n.a. R.E. Alekseev, 24 Minina Street, N. Novgorod 603950, Russia}

\date{\today}

\begin{abstract}
A method of windowed spatio-temporal spectral filtering is proposed to segregate different nonlinear wave components, and to calculate the surface of free waves. The dynamic kurtosis (i.e., produced by the free wave component) is shown able to contribute essentially to the abnormally large values of the surface displacement kurtosis, according to the direct numerical simulations of realistic sea waves. In this situation the free wave stochastic dynamics is strongly non-Gaussian, and the kinetic approach is inapplicable. Traces of coherent wave patterns are found in the Fourier transform of the directional irregular sea waves; they may form 'jets' in the Fourier domain which strongly violate the classic dispersion relation.	
%The values of dynamic  (i.e., caused by the free waves) kurtosis in typical sea states are calculated directly from the output data of phase-resolving numerical simulations of primitive water equations.
% with the help of a suggested method of wave decomposition. 
%The kurtosis remarkably exceeds the value peculiar to the Gaussian random process in rough sea states which are characterized by relatively narrow angle spectra. Then the free waves are found to make the major part of the total kurtosis.  This observation explicitly confirms inability of kinetic theories, which are based on the assumption of uncorrelated phases of free waves, to simulate such ‘abnormal' sea states. 
%
%The free waves are characterized by the value of kurtosis of the Gaussian statistics in other cases.
%
%The similarity of the conditions of these ‘abnormal’ sea states to the situation favorable to the Benjamin\-–Feir instability makes us believe that the revealed non-Gaussianity of free waves is caused by this physical mechanism. In accord, signatures of coherent wave patterns are observed in the Fourier transform of the directional irregular sea waves.
\end{abstract}
\maketitle

\section{Introduction}
Despite much efforts and remarkable progress in the study of the so-called rogue wave phenomenon \citep{Kharifetal2009,AkhmedievPelinovsky2010,Slunyaevetal2011,Onoratoetal2013,Dudleyetal2014,Akhmedievetal2016}, the consensus about their driving mechanisms in the open ocean has not been achieved yet (e.g. \citep{Residorietal2016} vs \citep{Fedeleetal2016}). The hydrodynamic wave processes in the limit of small nonlinearity and narrowband frequency and angle spectra may look much similar to signals in optical fibers, and then they are both governed by the nonlinear Schr\"{o}dinger equation (NLSE). The modulational instability caused by quasi-resonant four-wave interactions was suggested as a regular mechanism which increases the probability of large waves \cite{Onoratoetal2001}. The modulational instability of deep-water waves (the Benjamin--Feir instability, when talking in the hydrodynamic context) is captured by the focusing NLSE and its more accurate extensions. In particular, a number of exact solutions of the NLSE were proposed as rogue wave prototypes. Many of them have been successfully reproduced in hydrodynamic flumes and optical waveguides (see the recent review \citep{Dudleyetal2019}). 
%Today, solutions of other integrable equations, which possess qualitatively similar behavior, exhibiting a huge wave which appears from nowhere and disappears without trace \cite{Akhmedievetal2009} are frequently called 'rogue wave solutions' (e.g. ????). 

However, under realistic ocean conditions the efficiency of the Benjamin--Feir instability is disputed. On the one hand, besides theoretical grounds, the probability of high waves in experimental tanks was found much exceeding the conventional laws, when the wave conditions are favorable to the modulational instability \citep{Onoratoetal2009, Shemeretal2010}. On the other hand, results of processing of large amount of in-situ measurements presented in \citep{ChristouEwans2014, Fedeleetal2016} do not show noticeable deviation from the second-order statistical theory. Hence these works conclude that the Benjamin--Feir instability seemingly does not influence the real sea wave statistics.

So far the wave nonlinearity in operational probabilistic models is taken into account solely through the second-order Stokes wave corrections, which affect the wave shapes but do not play any role in the dynamical sense. In the Fourier representation the corresponding nonlinear components are commonly called phase-locked modes or \emph{bound waves}. \emph{Free waves} (the 'true' waves) are responsible for the natural wave modes of the dynamical system, they may evolve due to resonant or near-resonant interactions. This paradigm is fundamental for the present-day sea wave forecasting as it allows the use of phase-averaged kinetic equations, which can be simulated incomparably faster than the original equations of hydrodynamics. The non-Gaussianity of the free waves is totally neglected in the kinetic approach; in particular, the wave coherence necessary for the Benjamin--Feir instability is disregarded.    

The probabilistic wave properties are often estimated in terms of  statistical moments, which for the water surface $\eta$ are defined according to the relations
\begin{align} \label{StatisticalMoments}
	\sigma^2=\left< \eta^2 \right>, \quad
	\lambda_3=\frac{\left< \eta^3 \right>}{\sigma^3}, \quad
	\lambda_4=\frac{\left< \eta^4 \right>}{\sigma^4} -3,	
\end{align}
having zero mean, $\left< \eta \right>=0$. 
The moments of irregular deep-water waves in stationary conditions were estimated analytically in \citep{MoriJanssen2006} within the weakly nonlinear Hamiltonian theory.
The third statistical moment, skewness $\lambda_3$, is always small as its main contributor is proportional to the steepness $\epsilon\equiv k_p \sigma \ll 1$ ($k_p$ is the peak wavenumber), $\lambda_3 = 3\epsilon$. 
The fourth statistical moment, kurtosis $\lambda_4$, was shown to consist of two summands, $\lambda_4=\lambda_4^b + \lambda_4^d$.
The first, $\lambda_4^b = 24 \epsilon^2$, is due to bound waves, and is a small number. 
The other, dynamic kurtosis $\lambda_4^d = \pi/\sqrt{3}BFI^2$, is due to nonlinear coupling between free waves. Crucially, the Benjamin--Feir Index  $BFI=\sqrt{2}\epsilon/\Delta$ ($\Delta$ is the relative width of the frequency spectrum \citep{Onoratoetal2001,MoriJanssen2006}) is not necessarily small. The modulational instability may occur when $BFI>1$.  
Large values of kurtosis yield greater probability of large displacements and more frequent occurrence of high waves. Hence, the contribution of free wave non-Gaussian dynamics to the sea wave statistics theoretically may be even greater than the one from phase-locked modes.

Dynamical equations for the wave modulations (NLSE and its generalizations) are written in terms of the free wave complex amplitude, thus the dynamic kurtosis may be calculated straightforwardly within this approximate framework. Large deviation from the Gaussian statistics was observed in a number of direct numerical simulations of the NLSE-type models (e.g. \citep{Onoratoetal2001,Janssen2003,Socquetetal2005} and many others).
The dynamic kurtosis which exceeds the bound wave kurtosis was shown in the numerical simulations of unidirectional weakly nonlinear weakly modulated waves \citep{Shemeretal2010}, and within a more accurate framework, the Zakharov equations in canonical variables (i.e., the weakly nonlinear dynamic Hamiltonian theory) \citep{AnnenkovShrira2009}. In both the situations large values of $\lambda_4^d$ were registered during the fast transition stage. Recall that the theory \citep{MoriJanssen2006} was derived assuming stationary conditions; it was shown in \citep{SlunyaevSergeeva2011}, that the relation derived in \citep{MoriJanssen2006}  does not hold in such transition wave regimes. Furthermore, in \citep{AnnenkovShrira2018} the evolution of the wave kurtosis simulated by means of the primitive Euler equations, by the Zakharov equations and within the kinetic theory was shown to differ significantly. The remarkably better agreement was provided by the simulation of the primitive Euler equations. However, the integration of the primitive equation leads to the calculation of the total kurtosis; the distribution between the dynamic and bound kurtosis remains unknown.   

Hence, though it is generally accepted that under the conditions suitable for the Benjamin--Feir instability the probability of large waves should increase, the issue if coherent four-wave dynamics can influence significantly the wave statistics in the real ocean, remains unclear.   
The answer to this question is vital to the proper choice of the wave model capable of description of the rogue wave effect.  

In the present work we calculate the dynamic kurtosis directly from the data accumulated in phase-resolving simulations of the Euler equations. The considered irregular waves  possess the JONSWAP frequency spectrum, typical to the North Sea.  We show that under the conditions favorable for the modulational instability, the dynamic kurtosis attains large values similar in magnitude to the bound wave kurtosis, thus revealing strongly non-Gaussian properties of the free wave component. 
%The method of calculation of the free wave component does not employ any supplement assumptions.
We present the evidence of coherent wave patterns in the Fourier space, which are responsible for this effect. 

\section{Description of the method}
We solve the potential Euler equations for gravity waves, which propagate over infinitely deep water, with the help of the High Order Spectral Method \cite{Westetal1987}, accurately resolving four-wave nonlinear interactions.
%, which is one of the most efficient recognized algorithms. 
%In the literature, the wave data produced by this code have been verified against the laboratory measurements many times. 
%The code which takes into account up to the four-wave nonlinear wave interactions is used for the present study, which is capable of description of the Benjamin--Feir instability. 
%
The initial condition is specified in the form of a linear solution  with random Fourier phases (several realizations were simulated) and a prescribed JONSWAP shape of the averaged frequency spectrum with the given peak period $T_p = 10$~s, significant wave height $H_s$, peakedness factor $\gamma$ and the directional spread specified by the angle $\Theta$ (see details in \cite{SlunyaevKokorina2019}). 
The evolution of a wavy surface in the domain $50\times50$ dominant wave lengths
% (see in Fig.~\ref{fig:SurfaceSamples}) 
is simulated for $1200$ dominant wave periods, following $20$ wave periods of a preliminary stage when the  initially linear solution adiabatically adjusts to the nonlinear equations.  
%Only the cases were simulated when the wave breaking was not significant.
%, see other details of the simulation method in \cite{SlunyaevKokorina2019}.   

\begin{figure}
	%Figure
	%\centerline{\includegraphics[width=7.0cm]{MKdV-Soliton_Train2_Fig1_Snapshots.eps}(a)}
	\centerline{\includegraphics[width=7cm]{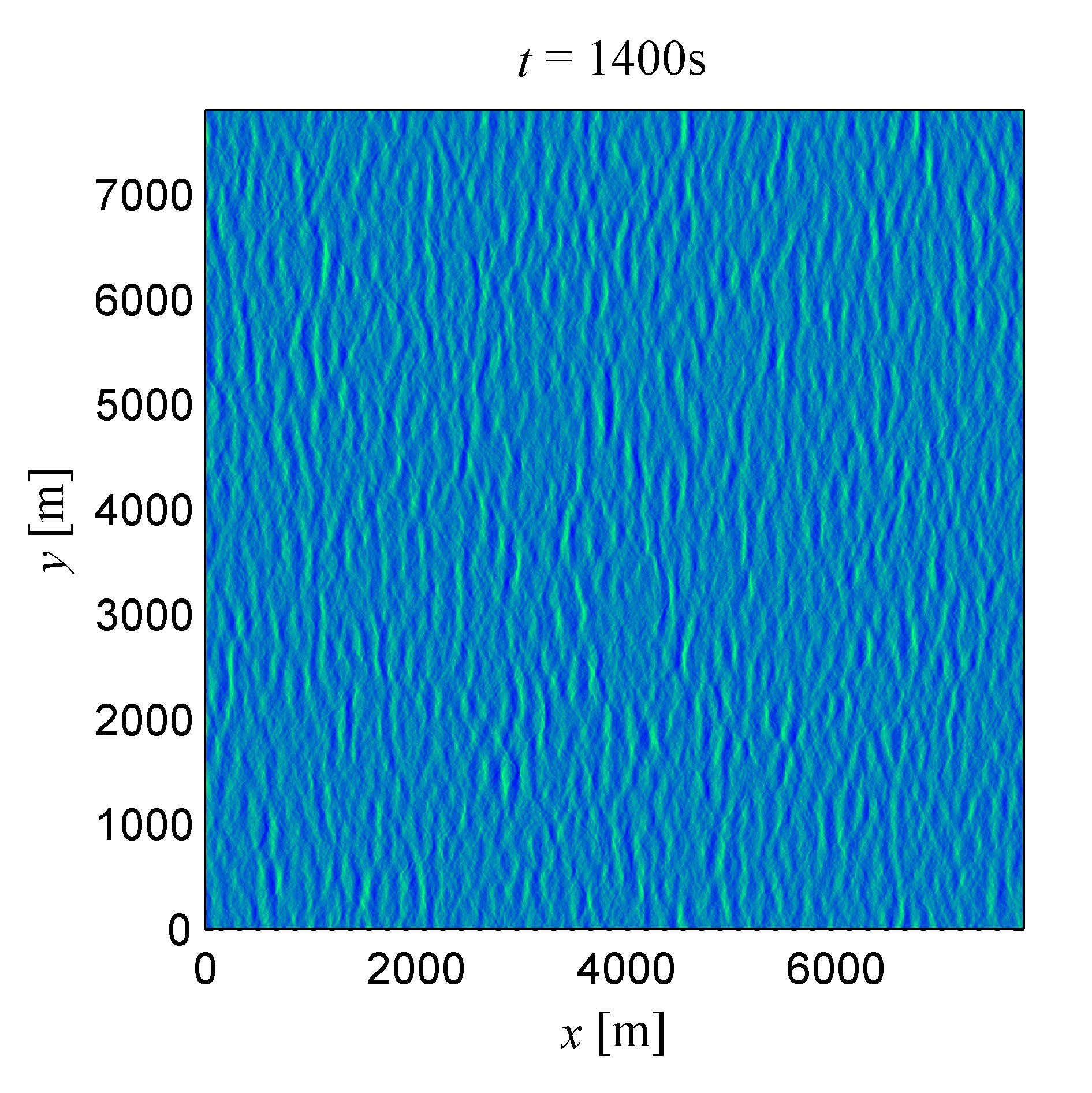}(a)}
	\centerline{\includegraphics[width=7cm]{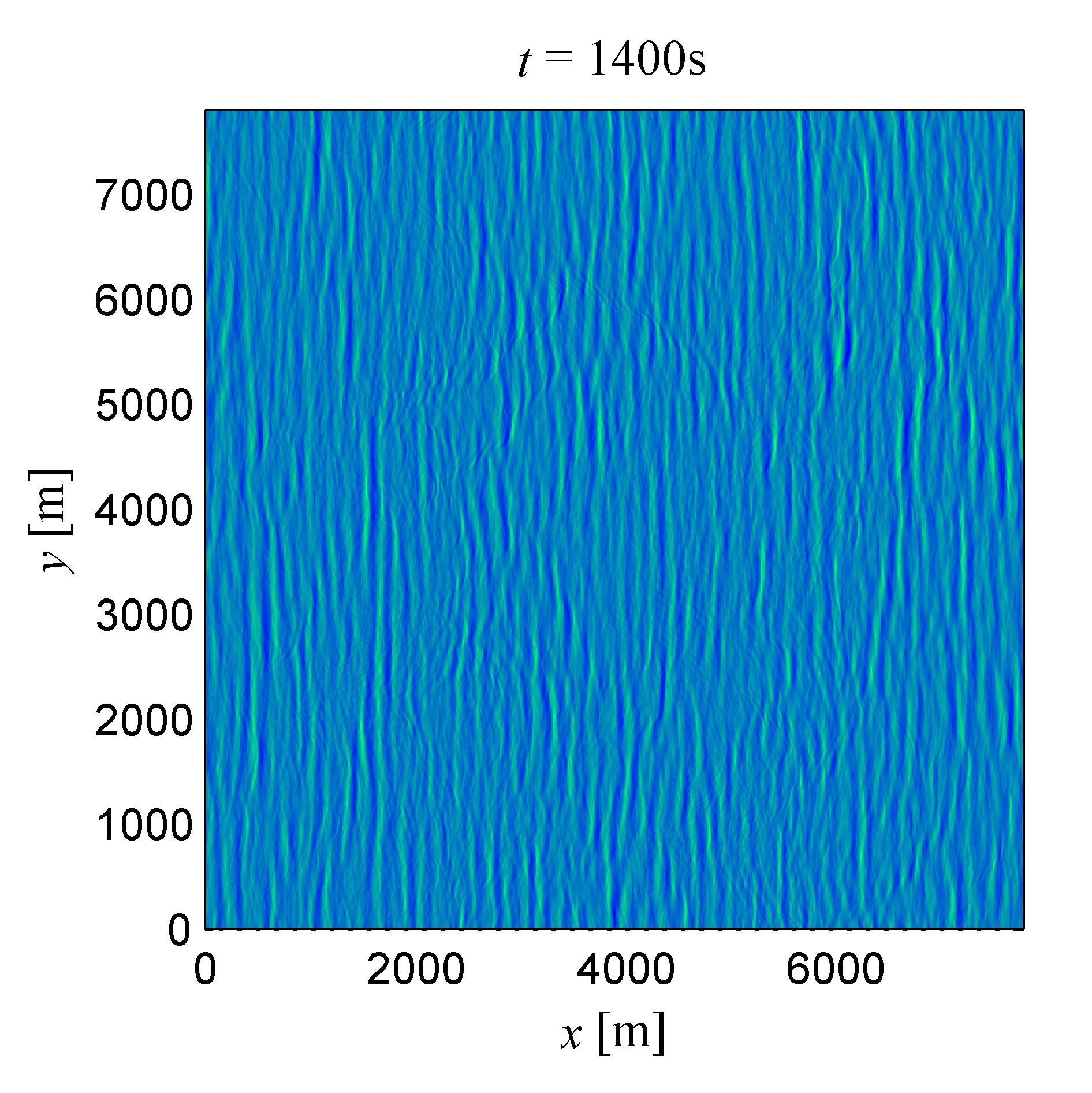}(b)}
	\caption{Instant water surfaces for the conditions $H_s = 7$~m, $\Theta = 62^\circ$, $\gamma=3$ (a), and $H_s = 6$~m, $\Theta = 12^\circ$, $\gamma=6$ (b) at the end of the simulations. }
	\label{fig:SurfaceSamples}
\end{figure}

The simulations represent evolution of sea waves in the overall area of about $500$ squared kilometers for $20$ minutes under given initial spectral conditions. The wind forcing is not included in the system; a weak hyperviscosity is introduced to the code to regularize occasional wave breaking events. Two samples of momentary surfaces are shown in Figs.~\ref{fig:SurfaceSamples}a,b for  the directional spreads $\Theta=62^\circ$ and $\Theta=12^\circ$; these two cases may be referred to as relatively short-crested and long-crested conditions respectively and will be considered further.  

\begin{figure}
	\centerline{\includegraphics[width=8.5cm]{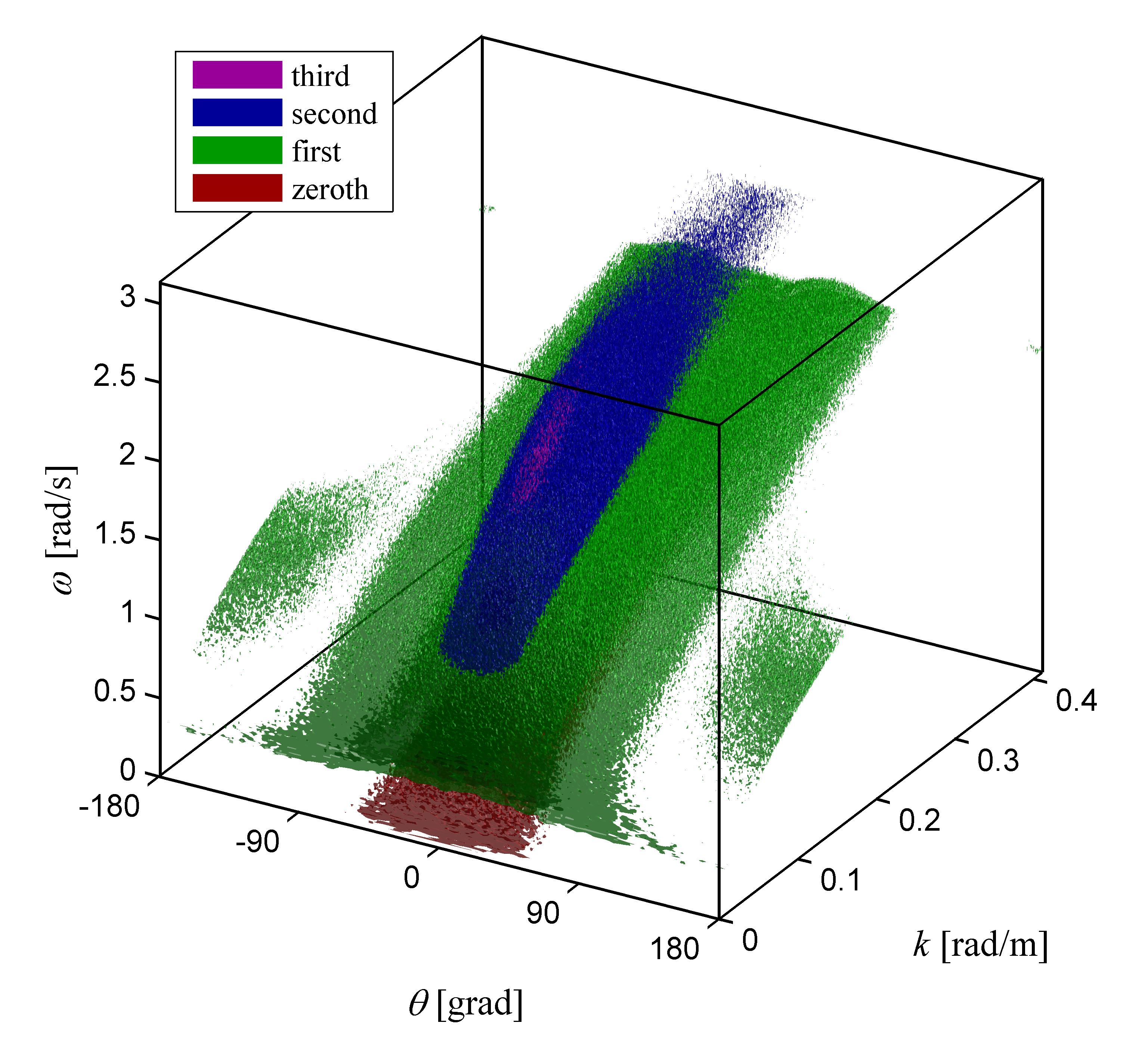}(a)}
	\centerline{\includegraphics[width=8.5cm]{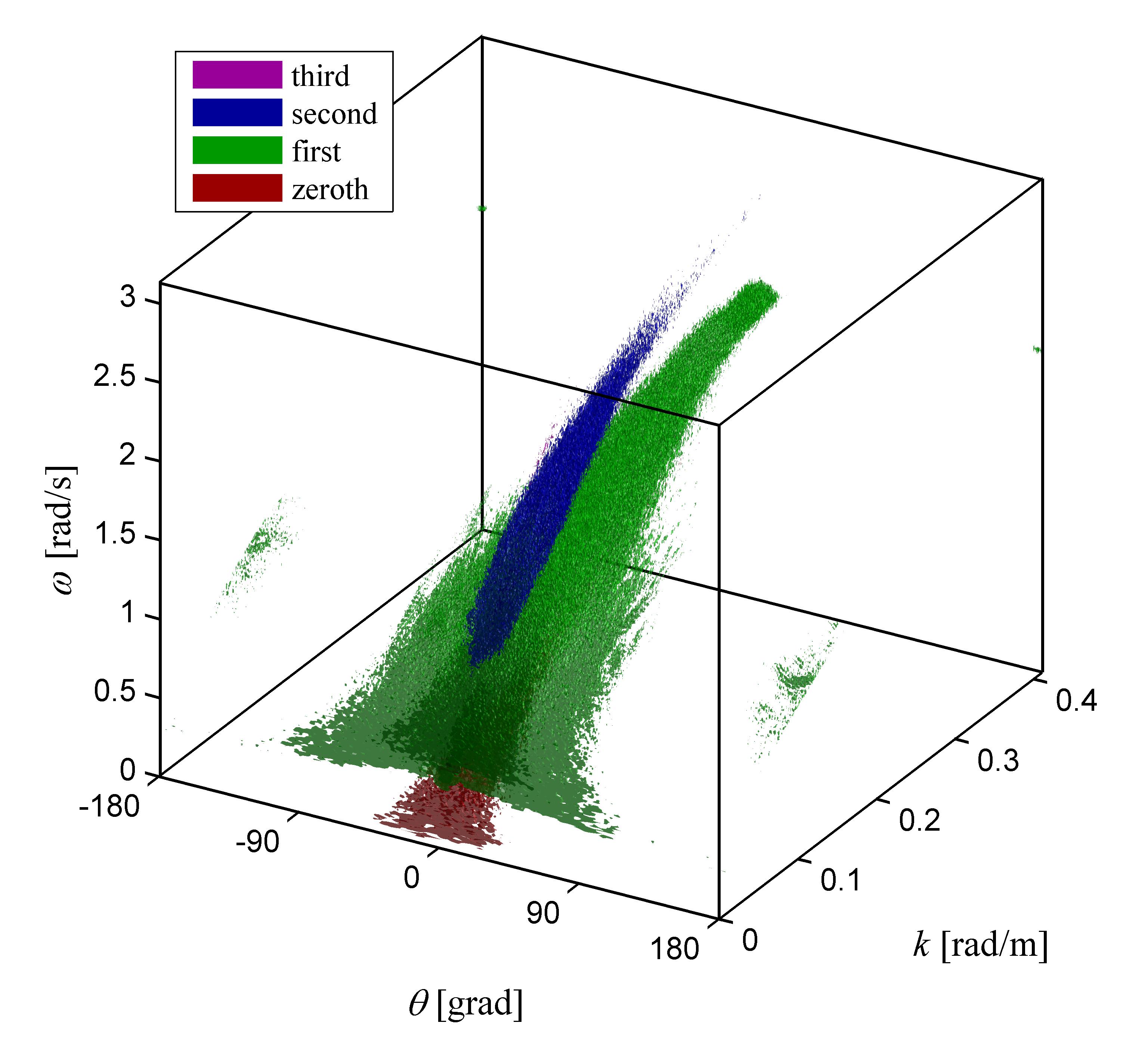}(b)}
	\caption{Isosurfaces of the Fourier amplitudes in the instant spatio-temporal transforms for the conditions $H_s = 7$~m, $\Theta = 62^\circ$, $\gamma=3$ (a) and $H_s = 6$~m, $\Theta = 12^\circ$, $\gamma=6$ (b) at the moments after more than 100 periods of the simulations. Different colors denote nonlinear harmonics. }
	\label{fig:Spectrum3D}
\end{figure}

\begin{figure}
	\centerline{\includegraphics[width=8.5cm]{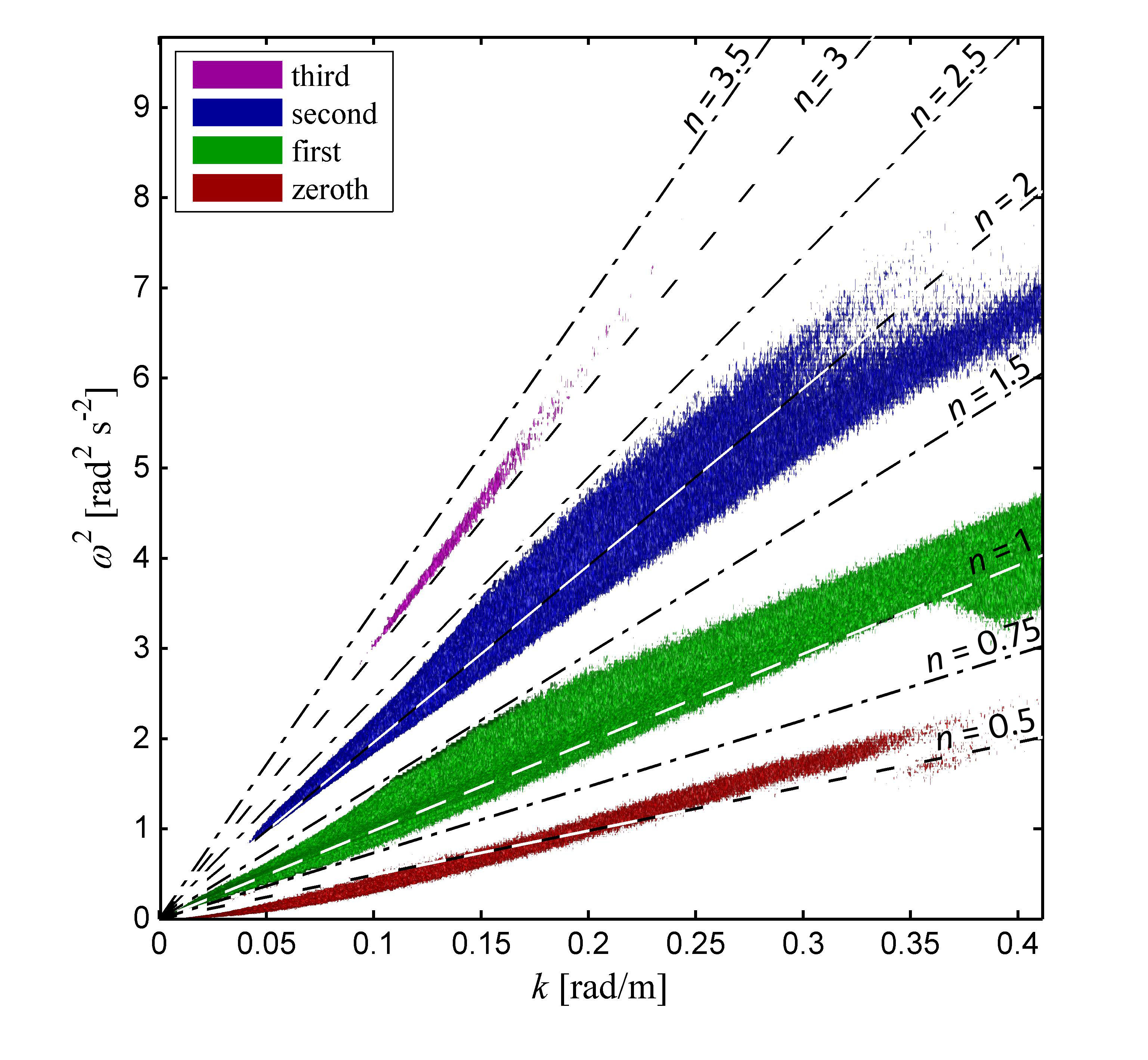}(a)}
	\centerline{\includegraphics[width=8.5cm]{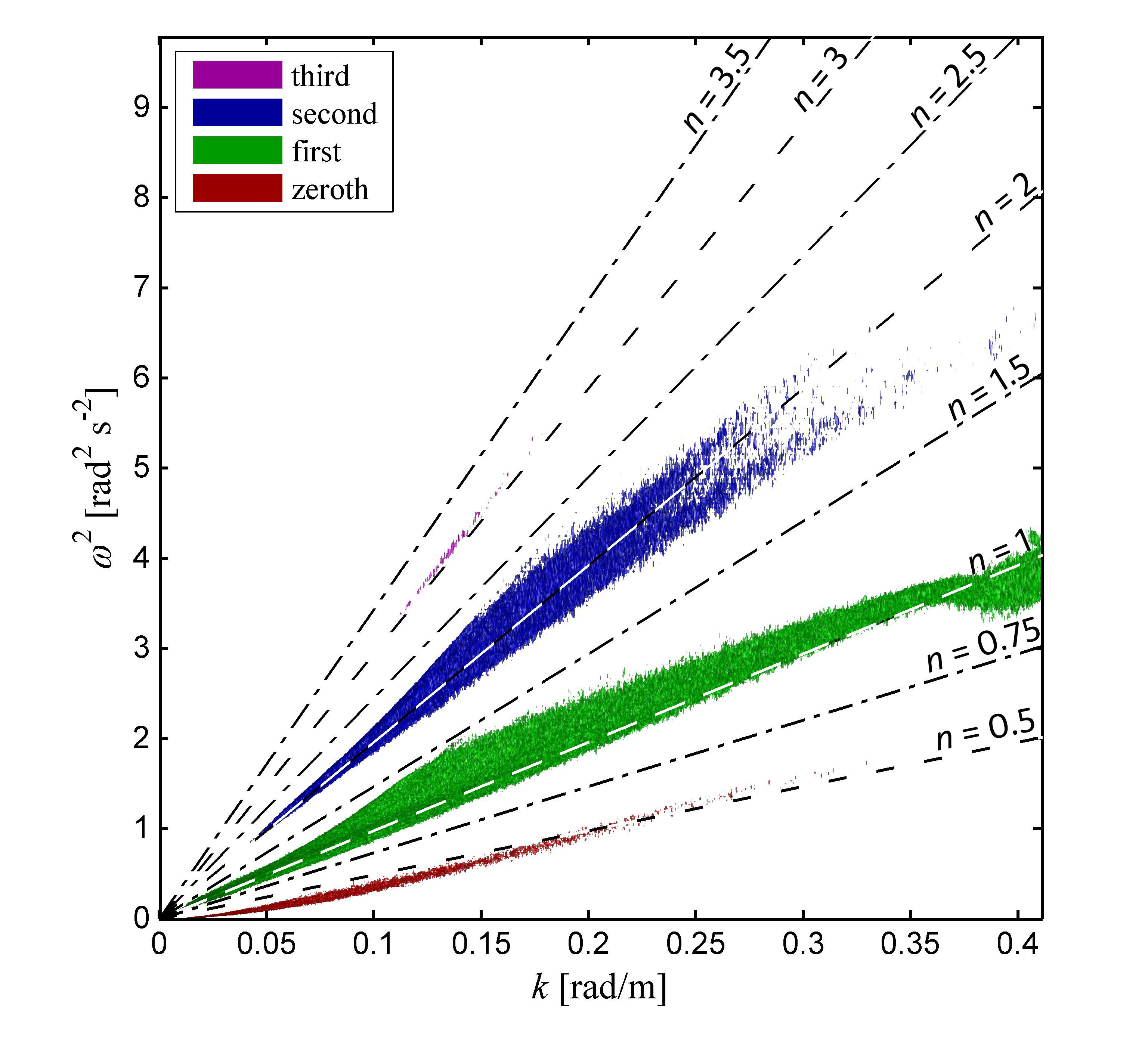}(b)}
	\caption{Same as in Fig.~\ref{fig:Spectrum3D}, but the view from the side along the $O\theta$ axis, and the vertical axis represents the squared frequency. The straight lines correspond to the relation (\ref{NonlinearHarmonics}) for different $n$. }
	\label{fig:Spectrum3DMarked}
\end{figure}

The method of evaluation of the free wave component from the surface displacement $\eta(x,y,t)$ operates in the Fourier domain. The instant Fourier amplitudes $\hat{\eta}$ are calculated after the application of the smoothing Hanning mask $M(t)=0.5 (1-\cos{2\pi t/T})$,  $T \approx 25 T_p$, which operates in time, and the Fourier transforms along the two coordinates and time,  
\begin{align} \label{Fourier}
%	\hat{\eta}(k_x,k_y,\omega,t) = \mathcal{F}_x \mathcal{F}_y \mathcal{F}_t \{ M(t-\tau) \eta(x,y,\tau) \}, \\ 
%	S(k,\theta,\omega,t)\equiv |\hat{\eta}(k_x,k_y,\omega,t)|.
	\hat{\eta}(k_x,k_y,\omega,t) = \mathcal{F}_x \mathcal{F}_y \mathcal{F}_\tau \{ M(\tau-t) \eta(x,y,\tau) \}.
%	, \\ 
%A(|k|,\theta,\omega,t) = |\hat{\eta}(k_x,k_y,\omega,t)|.
\end{align}
Here $k_x$, $k_y$ are components of the wave vector $\textbf{k}\equiv(k_x,k_y)$,  $\omega$ is the cyclic frequency. For our purpose it is convenient to represent the wave vector in polar coordinates with the absolute value  $k\equiv\sqrt{k_x^2+k_y^2}$ and the angle $\theta$ with respect to the dominant direction of the wave propagation along $Ox$, and to consider the function $A(k,\theta,\omega,t) = |\hat{\eta}|$. Then the quantity $kA^2$ has the meaning of momentary power spectral density in coordinates $k$, $\theta$ and $\omega$.

Examples of the spatio-temporal Fourier domains of short-crested and long-crested intense waves are shown in Fig.~\ref{fig:Spectrum3D} with the help of isosurfaces of the Fourier amplitudes $A$. Only the part $\omega>0$ is shown as the function $\eta$ is real-valued. The color intensity characterizes magnitudes of the Fourier amplitudes in the logarithmic scale, from the maximum down to $10^{-3}$ of it. Colors denote different nonlinear wave harmonics as will be discussed further. 

Assuming a narrow-banded process, dominant waves $(\textbf{k},\omega)$ through nonlinear $n$-order interactions generate Fourier harmonics $(n\textbf{k},n\omega)$, which in the weakly nonlinear approximation may be presented in the form
\begin{align} \label{NonlinearHarmonics}
	\omega_n = n \Omega(\textbf{k}/n), \quad \Omega(\textbf{k})\equiv \sqrt{gk},	
\end{align}
where $\omega_1=\sqrt{gk}$ is the dispersion relation, $g$ is the gravity acceleration, and integer $n\ge2$ is the order of interaction.   

Figs.~\ref{fig:Spectrum3DMarked}a,b give a better demonstration of the nonlinear harmonics. The figures are produced from the plots in Fig.~\ref{fig:Spectrum3D}a,b respectively, when they are rotated in such a way that the axis $O\theta$ gets perpendicular to the plane of the page. The vertical axis is transformed to the squared frequencies, then $\omega_n^2$ depend linearly on $k$, the corresponding lines are plotted in Fig.~(\ref{NonlinearHarmonics}).  
Remarkably, the calculated first ($n=1$), second ($n=2$) and third ($n=3$) harmonics of the broad JONSWAP spectrum are localized along the lines prescribed by the narrow-band relation (\ref{NonlinearHarmonics}). 

The 'difference' (or 'zeroth') harmonic produced by narrow-banded waves propagates with the group velocity of the carrier, and hence in the Fourier space corresponds to the condition $\omega = \textbf{C}_{gr}(\textbf{k}_c) \textbf{k}$, where $\textbf{C}_{gr}\equiv \nabla_{\textbf{k}} \Omega$ and $\textbf{k}_c$ is the carrier wave vector. In the conditions of broad-banded waves the location of the difference harmonic turns out to be relatively well described by the same condition (\ref{NonlinearHarmonics}) when $n=1/2$, see in Fig.~\ref{fig:Spectrum3DMarked} (this should correspond to the decay interaction). Significantly, the use of (\ref{NonlinearHarmonics})  does not require an assignment of $\textbf{k}_c$. 

Thus, the first, second, third and the 'zeroth' nonlinear harmonics (the latter is the induced long-scale displacement) may be easily distinguished.
What is essential, the corresponding volumes in the Fourier space almost do not intersect in the situations of narrow and relatively broad angle spectra shown in Fig.~\ref{fig:Spectrum3DMarked}. More exactly, the corresponding overlapping parts of $|\hat{\eta}(k_x,k_y,\omega)|$ contain tiny amount of energy. 
Hence, the nonlinear harmonics may be segregated with the help of a simple spectral filter. In Fig.~\ref{fig:Spectrum3DMarked} the dash-dotted lines show the conditions (\ref{NonlinearHarmonics}) for $n=0.75$, $1.5$, $2.5$, which specify the boundaries between the nonlinear harmonics. The color coding in Fig.~\ref{fig:Spectrum3D} and Fig.~\ref{fig:Spectrum3DMarked} denotes the selected in this way harmonics. It may be seen from Fig.~\ref{fig:Spectrum3DMarked} that such a simple filter manages very well in both the cases (the process of spectral segregation may be further improved if necessary).
%An example of relatively long-crested intense waves is presented below in Fig.~\ref{fig:CoherentPatterns}. 
%In general, the first and zeroth harmonics are selected particularly accurately.

To select the first harmonic, we introduce the spectral mask $N(k_x,k_y,\omega)$ which is equal to one if $\omega_{0.75}(k) < \omega < \omega_{1.5}(k)$ and is zero otherwise (here $\omega_{0.75}$ and $\omega_{1.5}$ are determined by (\ref{NonlinearHarmonics})). Then the first harmonic $\eta_{1}$  is calculated after  the inverse to (\ref{Fourier}) triple Fourier transform,   
\begin{align} \label{InverseFourier}
\eta_{1}(x,y,t) =\mathcal{F}_x^{-1} \mathcal{F}_y^{-1} \mathcal{F}_\tau^{-1} \{ N \hat{\eta} \}.
\end{align}
%
%Note that the instant $t$ is related to the moment of maximum of the Hanning mask, when $M=1$. 
%
Strictly speaking the first harmonic contains a constituent of phase-locked waves, but its magnitude is at least two orders of steepness smaller than the amplitude of free waves. Therefore, in what follows we will assume that the first harmonic and the free wave component are equivalent.

We note that waves which travel opposite to the principal wave direction may be seen in Fig.~\ref{fig:Spectrum3D} (the angles $|\theta|>90^\circ$). They did not exist originally, but reach up to about $0.15\%$ of the total wave energy for the 20 minutes of simulation of the short-crested waves (Fig.~\ref{fig:Spectrum3D}a). The portion of opposite waves in the long-crested wave field is smaller. The generation of opposite waves was observed in precise numerical simulations of initially unidirectional waves in the planar geometry in  \citep{Slunyaev2018}. 
The distribution of very short waves in Fig.~\ref{fig:Spectrum3DMarked} is somehow affected by the weak hyperviscosity which was introduced to suppress breaking of too steep waves. 

The frequency-wavenumber plots were used in previous works to separate spectral areas responsible for different nonlinear harmonics (e.g., \citep{KrogstadTrulsen2010,Takloetal2015,Takloetal2017,Slunyaev2018}). We emphasize that in the present study we make use of the available Fourier phases, and reconstruct the surfaces of free and bound waves. This approach opens new possibilities such as evaluation of the statistical characteristics of the wave components directly, with no extra assumptions.  

\section{Full and dynamic statistical moments}
Instant full statistical moments of the water surface $\eta(x,y)$ are calculated according to (\ref{StatisticalMoments}), where the angle brackets denote the averaging over the surface in one realization. Different realizations are used to estimate the confidence interval. The dynamic statistical moments at the given time instant are calculated similarly, based on the obtained free wave surfaces  $\eta_1(x,y)$ (\ref{InverseFourier}).

\begin{figure}
	%Figure
	\centerline{\includegraphics[width=8.5cm]{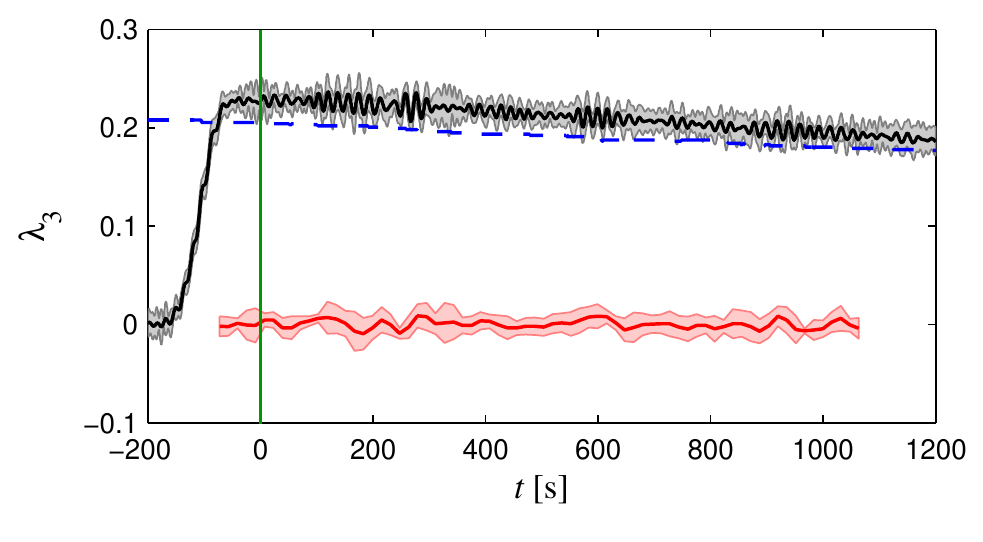}(a)}
	\centerline{\includegraphics[width=8.5cm]{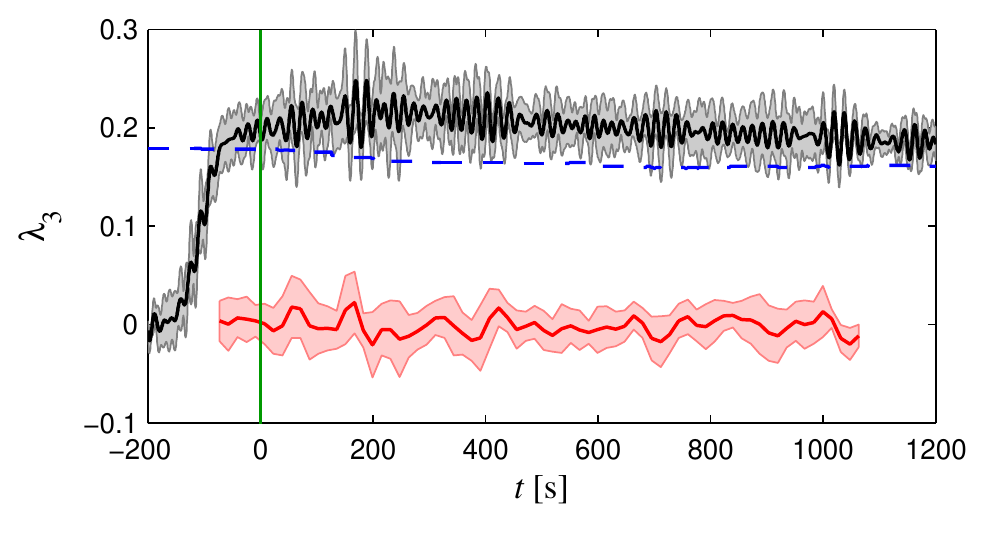}(b)}
	\caption{Skewness in the sea states with the parameters $H_s = 7$~m, $\Theta = 62^\circ$, $\gamma=3$ (a), and $H_s = 6$~m, $\Theta = 12^\circ$, $\gamma=6$ (b). The black thick curve and gray shading (above) show the ensemble averaged total skewness and its standard deviation band. The red line and shading below show the dynamic part of the skewness. The broken lines give the skewness estimation $\lambda_3=3\epsilon$.}
	\label{fig:Skewness}
\end{figure}
\begin{figure}
	%Figure
	\centerline{\includegraphics[width=8.5cm]{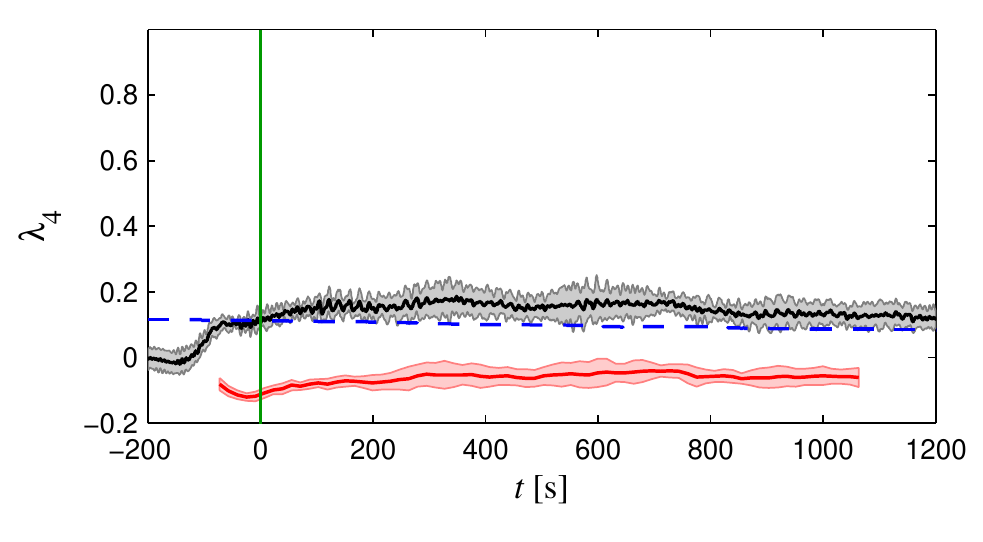}(a)}
	\centerline{\includegraphics[width=8.5cm]{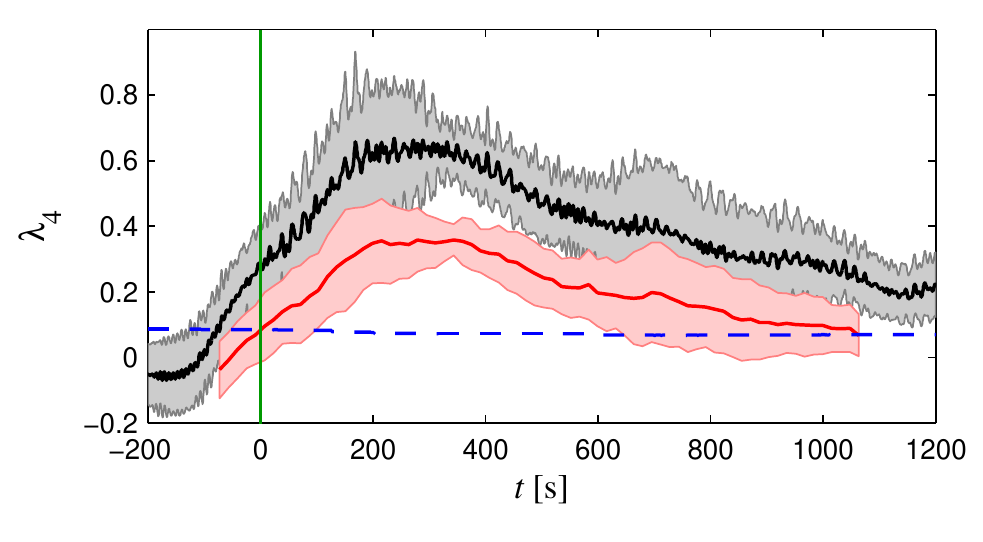}(b)}
	\caption{Kurtosis for the same parameters as in Fig.~\ref{fig:Skewness}. The total kurtosis is above and the dynamic kurtosis is below. The broken lines give the estimation of the bound wave kurtosis $\lambda_4^b=24\epsilon^2$.}
	\label{fig:Kurtosis}
\end{figure}

The full and dynamic variances $\sigma$ remain approximately constant unchanged throughout the period of simulation, as the evolution is almost conservative. Meanwhile the asymmetry $\lambda_3$ and the kurtosis $\lambda_4$ are time dependent. They are shown in Fig.~\ref{fig:Skewness} and Fig.~\ref{fig:Kurtosis} for the two representative cases. Above, the black lines and gray shading represent the full statistical moments with the confidence interval of one standard deviation. The lower red curves with shaded areas give the dynamic parts of the skewness and kurtosis. As was mentioned earlier, the first $200$ seconds of the simulations are preparatory and should not be considered (left to the vertical lines in Fig.~\ref{fig:Skewness} and Fig.~\ref{fig:Kurtosis}, $t<0$).

The plots of the wave surface skewness look very similar for the two situations displayed in Fig.~\ref{fig:Skewness}. The dynamic skewness is about zero; the total skewness possesses similar values in these two cases, having a trend to decay gradually due to slow broadening of the angle spectrum. The analytic estimations of $\lambda_3$ according to \cite{MoriJanssen2006} are shown with the broken lines; they underestimate the actual values a little. The skewness in Fig.~\ref{fig:Skewness}b exhibits somewhat larger spread.     

The situation with the kurtosis is qualitatively different. The total kurtosis attains much larger values in the simulation of steep long-crested waves (Fig.~\ref{fig:Kurtosis}b). In this case it exhibits a rapid growth and subsequent slow decrease in agreement with numerous previous studies (e.g., \cite{Janssen2003,Socquetetal2005,Onoratoetal2009,Shemeretal2010,AnnenkovShrira2009,SlunyaevSergeeva2011,AnnenkovShrira2018}), while short-crested waves are characterized by a smaller value of $\lambda_4$ which almost does not vary in time and is close to the estimation of the bound wave kurtosis $\lambda_4^b$ obtained in \cite{MoriJanssen2006}. We do not plot the curves for the dynamic kurtosis $\lambda_4^d$ since the method for calculation of $BFI$ is ambiguous. 

The dynamic kurtosis is steady in Fig.~\ref{fig:Kurtosis}a for short-crested waves; it is slightly below zero, what may be explained by the finite statistical ensemble. 
The dynamic kurtosis of long-crested waves evolves qualitatively similar to the full kurtosis. Importantly, it makes up about one half of the total kurtosis, by about 0.3 in the absolute value.
During the transition stage the theoretical bound wave kurtosis $\lambda_4^b$ (the broken line in Fig.~\ref{fig:Kurtosis}b) lies much below the dynamic kurtosis of simulated waves.     

The essential departure of the dynamic kurtosis from zero reveals the non-Gaussian wave dynamics. In contrast to the previous studies, we show explicitly within the primitive equations of hydrodynamics the strong non-Gaussianity of free waves, which are the 'normal modes' of the wave system.

\section{'Signatures' of coherent wave patterns}
It was shown in previous studies, that under the conditions of intense irregular unidirectional waves with a narrow spectrum, wave groups of the shape similar to NLSE envelope solitons can appear (e.g. \citep{Slunyaev2006,Viottietal2013}), however if the waves were indeed coherent (i.e., the groups persist) was unclear.
To investigate this issue, we examine the evolution of the Fourier amplitudes for the situation of intense long-crested waves.

\begin{figure}
	%Figure
	\centerline{\includegraphics[width=9.0cm]{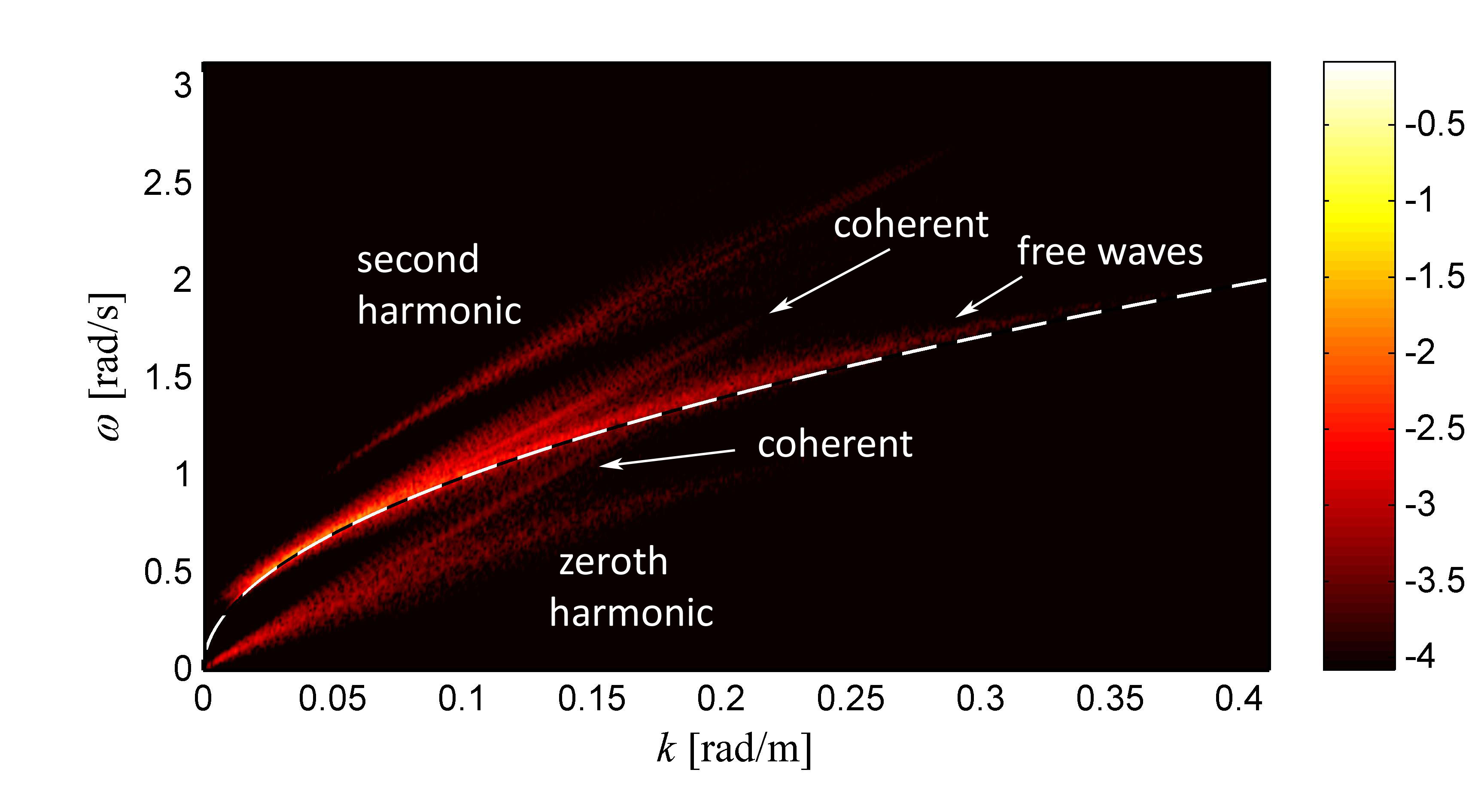}(a)}
	\centerline{\includegraphics[width=9.0cm]{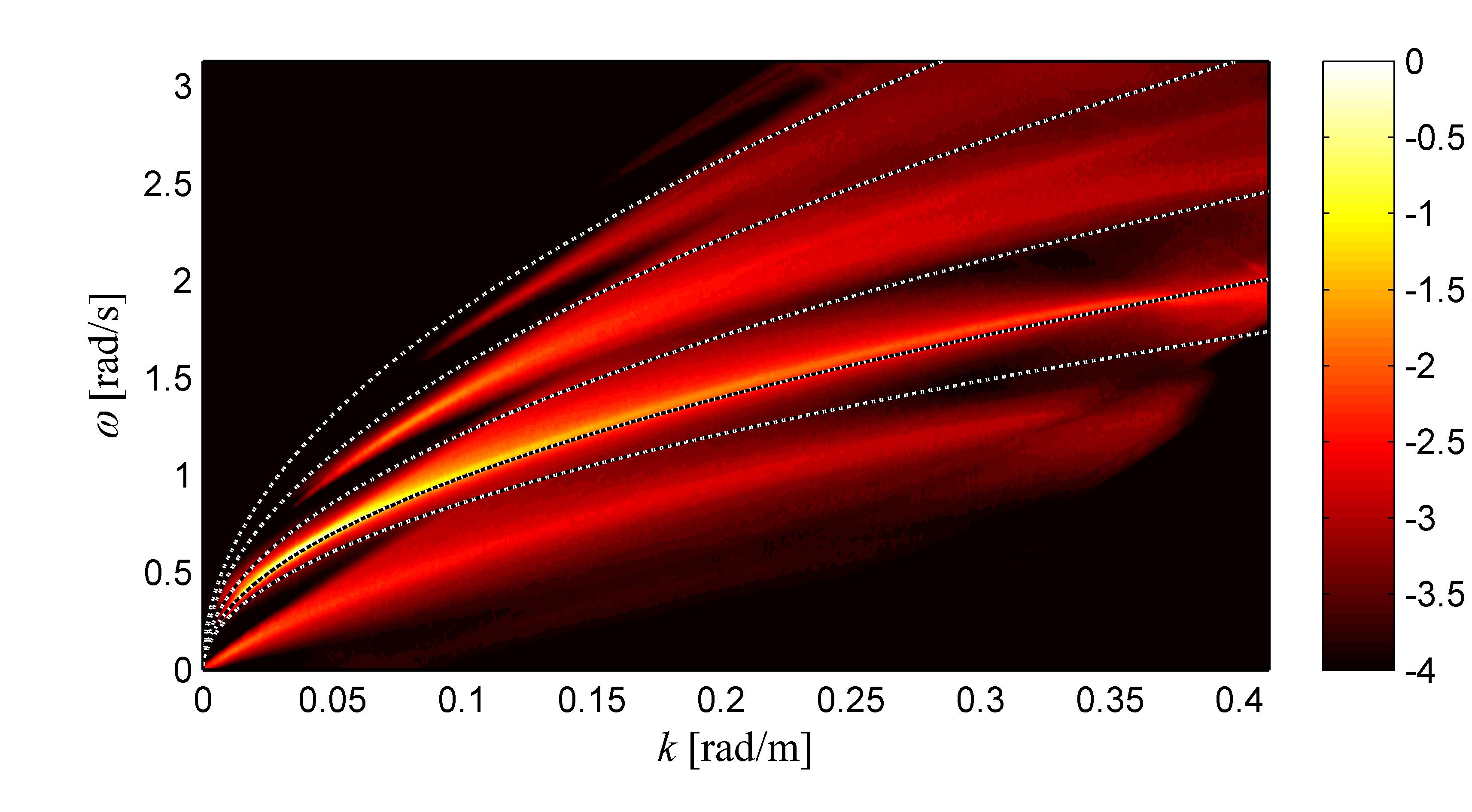}(b)}
	\caption{The section of Fourier amplitudes $A(k,\theta,\omega,t)$ for  $\theta \approx 14^{\circ}$ and $t \approx 45T_p$ (a), and the amplitude spectral density when integrated over the angle, $\sqrt{ k \int{A^2 d\theta}}$ (b). The sea state conditions are $H_s = 6$~m, $\Theta = 12^\circ$ and $\gamma=6$. The dashed lines show the linear dispersion law and the bands of different nonlinear harmonics according to the spectral filter.}
	\label{fig:CoherentPatterns}
\end{figure}

An example of the instant Fourier transform for the given angle $\theta \approx 14^{\circ}$ is shown in Fig.~\ref{fig:CoherentPatterns}a. The free wave component follows the dispersion curve (shown by the broken line); it is slightly above due to the nonlinear frequency shift. Besides, one may clearly see a remarkable violation of the dispersion law: significant amount of energy is distributed along the tangent line. This spectral 'jet' corresponds to coherent wave structures (one may discern many) which travel with speeds approximately equal to the group velocity. These structures are supported by quasi-resonant and non-resonant interactions between the dominant and close side harmonics (a discussion of similar Fourier portraits observed in fully nonlinear simulations of unidirectional narrow-banded waves may be found in \citep{Slunyaev2018}; they may be also found in the simulations of the NLS equation in \cite{KrogstadTrulsen2010}). In the present example the second nonlinear harmonic (straight elongated area above) is determined mainly by the coherent part of the free waves. The difference nonlinear harmonic consists of two distinctive lobes (shown with arrows below the dispersion curve), which result from the coherent and incoherent parts of the first harmonic.     

When the wave energy is integrated over all angles $\theta$, the lobes which represent the coherent patterns cannot be seen (Fig.~\ref{fig:CoherentPatterns}b). All together they lead to broadening of the spectral lobes in addition to the nonlinear frequency upshift, and form the actual lobes of nonlinear harmonics. 

\section{Conclusion}
In this communication we show explicitly by virtue of the direct numerical simulation of primitive equations of hydrodynamics, that irregular nonlinear directional sea waves may violate essentially the Gaussian statistics due to the coherent dynamics of free waves. Under the condition of relatively narrow angle spectrum the dynamic kurtosis may be comparable with the value of the bound wave kurtosis. The method which allows segregation of different nonlinear harmonics, including the free wave component, in broad-banded directional deep-water waves is suggested and employed.

The coherent 3D wave patterns persist in the irregular directional sea wave fields and do not follow the classic dispersion relation. Consequently, they noticeably contribute to the spread of the actual relation between wave vectors and frequencies. Still, the lobes of the nonlinear harmonics in the spatio-temporal Fourier domain overlap weakly, what enables their efficient separation with the help of the suggested method.  

In our opinion, the contradicting statements about the role of the Benjamin--Feir instability in the real sea, mentioned in the introduction, may be brought together if the situations when the waves suffer from the Benjamin--Feir instability are not frequent. The strong deviation from the conventional wave height probability in these abnormal sea states will be inconspicuous when averaged over a much greater amount of 'ordinary' sea states. This circumstance, however, does not mean that the situations of 'abnormal' wave statistics caused by strong wave coherence are not significant for the ocean.    

%\begin{acknowledgment}
The study is supported by the RSF grant No 16-17-00041.
%\end{acknowledgment}

\end{document}